# On the rheology of a liquid-vapor interface

Roumen Tsekov and Boryan Radoev
Department of Physical Chemistry, University of Sofia, 1164 Sofia, Bulgaria

The mass and momentum balances are theoretically studied in heterogeneous two-component systems. Following Gibbs the system is presented as two bulk and a single surface phases. Comparing the equations derived with some typical rheological models, useful information about the location of the interface is obtained. It was demonstrated that the surface phase for insoluble surfactants coincides with the equimolecular interface, while for soluble ones it is placed on the surface of total mass density zero excess. In both cases the surface phase is close to the surface of tension and kinematic surface.

In spite of the doubtless progress made in the understanding of transfer processes in heterogeneous systems, there is no rigorous derivation of the phenomenological equations governing mass, momentum and energy balances. The microscopic theories, dealing with complicate mathematical formalism, provide description based on first principles [1]. However, it is not possible till now to translate their results in the useful language of macroscopic physics. The methods of non-equilibrium thermodynamics [2, 3] and hydrodynamics [4, 5] are easier to handle and seem to be more appropriate for description of the problem. After Gibbs the heterogeneous systems are described as homogeneous bulk and surface phases interchanging matter, momentum, energy, etc. In the literature [6-8] the surface balances are usually presented in analogy with the bulk ones. An important point here is the determination of the location of surface phases. There are number of dividing surfaces [2] corresponding to different conservation laws (surface of tension, equimolecular surface, etc.) where the surface phase could be placed. The goal of the present paper is to show the location of the surface phases in some typical cases on the base of existing experimental investigations of their rheology.

Let us consider a macroscopic system consisting of two bulk phases, liquid and vapor. The mass density of the liquid denoted by $\rho$ is much higher than the vapor mass density and for this reason we accept the latter equal to zero. Let the considered liquid be a two-component solution with mass concentration $c$ of the second component and mass concentration $\rho - c$ of the solvent. The balances in the bulk obey the mass and momentum conservation laws [9]

$$\partial_t \rho + \nabla \cdot (\rho \mathbf{v}) = 0 \tag{1a}$$

$$\partial_t c + \nabla \cdot (c\mathbf{v} + \mathbf{J}) = 0 \tag{1b}$$

$$\partial_t (\rho \mathbf{v}) + \nabla \cdot (\rho \mathbf{v}\mathbf{v} - \mathbb{P}) = 0 \tag{1c}$$

where $\mathbf{v}$ is the hydrodynamic velocity, $\mathbf{J}$ is the diffusion flux, $\mathbb{P}$ is the stress tensor and $\nabla$ is the nabla operator. The flux $\mathbf{J}$ can be expressed by the Fick law expressed properly by the mass fraction [9]

$$\mathbf{J} = -\rho D \nabla (c/\rho) \qquad (2a)$$

In the case of a Newtonian liquid the stress tensor takes the form

$$\mathbb{P} = -p\mathbb{I} + \rho\nu[\nabla\mathbf{v} + (\nabla\mathbf{v})^{\mathrm{T}} - (2/3)(\nabla\cdot\mathbf{v})\mathbb{I}] + \rho\mu(\nabla\cdot\mathbf{v})\mathbb{I} \qquad (2b)$$

where $\mathbb{I}$ is the unit tensor and the superscript $^{\mathrm{T}}$ denotes the conjugated tensor. In general the diffusion coefficient $D$ and kinematic shear $\nu$ and dilatational $\mu$ viscosities are functions of the mass density $\rho$, concentration $c$ and temperature $T$. Assuming a local equilibrium in the liquid the pressure can also be expressed as a function of these quantities via the corresponding equation of state $p(\rho,c,T)$.

According to the Gibbs concept, the real system under consideration is equivalent to two homogeneous bulk phases, liquid and vapor, divided by a two-dimensional homogeneous surface phase. Therefore, the laws of conservation at the surface should have the same form as eqs. (1) with non-zero right hand sides accounting for the interaction with the bulk [5, 10]

$$\partial_t \Gamma_{tot} + \nabla_\tau \cdot (\Gamma_{tot}\mathbf{u}) = [\rho(\mathbf{v}-\mathbf{u})]\cdot\mathbf{n} \qquad (3a)$$
$$\partial_t \Gamma + \nabla_\tau \cdot (\Gamma\mathbf{u} + \mathbf{J}_S) = [c(\mathbf{v}-\mathbf{u}) + \mathbf{J}]\cdot\mathbf{n} \qquad (3b)$$
$$\partial_t (\Gamma_{tot}\mathbf{u}) + \nabla_\tau \cdot (\Gamma_{tot}\mathbf{u}\mathbf{u} - \mathbb{P}_S) = [\rho\mathbf{v}(\mathbf{v}-\mathbf{u}) - \mathbb{P}]\cdot\mathbf{n} \qquad (3c)$$

Here $\Gamma_{tot}$ and $\Gamma$ are the total and second component surface excess mass densities, $\mathbf{u}$ is the hydrodynamic velocity of the surface phase, $\mathbf{J}_S$ and $\mathbb{P}_S$ are the interfacial diffusion flux and stress tensor, $\mathbf{n}$ is the unit normal vector to the surface, $\nabla_\tau \equiv \mathbb{U}\cdot\nabla$ and $\mathbb{U} \equiv \mathbb{I} - \mathbf{nn}$ are the surface nabla operator and tangential unite tensor. The right hand side of eqs. (3) represent the bulk mass and momentum fluxes to the interface.

The unit normal vector obeys the following equation [3, 10]

$$\partial_t \mathbf{n} + \nabla_\tau (\mathbf{n}\cdot\mathbf{u}) = 0 \qquad (4a)$$

Since the interface is a separate two-dimensional phase, the diffusion flux $\mathbf{J}_S$ and stress tensor $\mathbb{P}_S$ should be similar to the bulk ones. Hence, the Fick law (2a) for the surface phase reads

$$\mathbf{J}_S = -\Gamma_{tot} D_S \nabla_\tau (\Gamma / \Gamma_{tot}) \tag{4b}$$

where $D_S$ is the surface diffusion coefficient. In the case of a Newtonian interfacial fluid, the surface stress tensor $\mathbb{P}_S$ can be in analogy to eq. (2b) presented as

$$\mathbb{P}_S = -\pi \mathbb{U} + \Gamma_{tot} \nu_S [(\nabla_\tau \mathbf{u}) \cdot \mathbb{U} + \mathbb{U} \cdot (\nabla_\tau \mathbf{u})^T - (\nabla_\tau \cdot \mathbf{u}) \mathbb{U}] + \Gamma_{tot} \mu_S (\nabla_\tau \cdot \mathbf{u}) \mathbb{U} + \Gamma_{tot} \chi_S (\nabla_\tau \mathbf{u}) \cdot \mathbf{nn} \tag{4c}$$

where $\pi$ is the two dimensional pressure, $\nu_S$ and $\mu_S$ are the surface shear and dilatational kinematic viscosities. The last term in eq. (4c) accounts for the transverse shear flow [8, 11] with corresponding kinematic viscosity $\chi_S$. It is reasonable, the surface diffusion coefficient and kinematic viscosities, having the same dimension with the bulk transport coefficients, to depend on the excess mass densities and temperature. The presentation of the dynamic viscosities in eq. (4c) as products of the total mass excess density $\Gamma_{tot}$ and the corresponding kinematic viscosities is similar to the bulk one and accounts for the necessary condition the total mass excess and dynamic viscosities to have one and the same sign (remember that the mass excess densities could be either positive or negative).

The general description of the coupled bulk and interfacial transports via eqs. (1-4) is too complicated. A good simplification is the model of incompressible liquids. Further, in the case of small deviations from the equilibrium the non-linear terms in the balances of conservation drop out, eqs. (1-2) reduce to

$$\nabla \cdot \mathbf{v} = 0 \tag{5a}$$
$$\partial_t c = D \Delta c \tag{5b}$$
$$\partial_t \mathbf{v} = -\nabla p / \rho + \nu \Delta \mathbf{v} \tag{5c}$$

and eqs. (3-4) get the forms

$$\partial_t \Gamma_{tot} + \Gamma_{tot} \nabla_\tau \cdot \mathbf{u} = \rho (v_n - u_n) \tag{6a}$$
$$\partial_t \Gamma + \Gamma \nabla_\tau \cdot \mathbf{u} = D_S [\Delta_\tau \Gamma - (\Gamma / \Gamma_{tot}) \Delta_\tau \Gamma_{tot}] + c(v_n - u_n) - D \nabla_n c \tag{6b}$$
$$\Gamma_{tot} \partial_t \mathbf{u}_\tau = -\nabla_\tau \pi + \Gamma_{tot} \nu_S \Delta_\tau \mathbf{u}_\tau + \Gamma_{tot} \mu_S \nabla_\tau (\nabla_\tau \cdot \mathbf{u}) - \rho \nu (\nabla_\tau v_n + \nabla_n \mathbf{v}_\tau) \tag{6c}$$
$$\Gamma_{tot} \partial_t u_n = \pi \nabla_\tau \cdot \mathbf{n} + \Gamma_{tot} \chi_S \Delta_\tau u_n + p - 2\rho \nu \nabla_n v_n \tag{6d}$$
$$\partial_t \mathbf{n} + \nabla_\tau u_n = 0 \tag{6e}$$

where $\Delta_\tau \equiv \nabla_\tau \cdot \nabla_\tau$ and $\mathbf{u}_\tau = \mathbb{U} \cdot \mathbf{u}$. Equations (6) are the boundary conditions to eqs. (5).

Let us consider first the case of a pure solvent, i.e. $c=0$ and $\Gamma=0$. In this case the boundary conditions to eqs. (5) are usually described by the following equations [5]

$$\nabla_\tau v_n + \nabla_n \mathbf{v}_\tau = 0 \qquad \sigma\Delta_\tau \zeta + p - 2\rho\nu\nabla_n v_n = 0 \qquad \partial_t \zeta = v_n$$

where $\sigma$ and $\zeta$ are the thermodynamic surface tension and the normal deformation of the surface, respectively. These equations will coincide with eqs. (6) if $\pi = -\sigma$, $\mathbf{u} = \mathbf{v}$ and $\Gamma_{tot} = 0$. Therefore, the surface phase in one-component systems should be placed at the surface of tension determined by the elastic force and momentum balances, kinematic surface, where the bulk and surface velocities are equal, and equimolecular surface, where the surface mass density excess $\Gamma_{tot} - \Gamma$ of the solvent is zero. The rigorous calculations [2] show that these three surfaces do not coincide in general. However, the differences are too small (in order of few angstroms) and for this reason they are not important for the macroscopic hydrodynamics [4].

For the two-component systems, there are two distinguished models of the surface rheology reported in the literature. The first one is usually applied for insoluble monolayers ($c=0$). This case is well described by the following boundary conditions

$$\partial_t \Gamma + \Gamma \nabla_\tau \cdot \mathbf{v} = 0 \tag{7a}$$
$$\Gamma \partial_t \mathbf{v}_\tau = \partial_\Gamma \sigma \nabla_\tau \Gamma + \Gamma \nu_S \Delta_\tau \mathbf{v}_\tau + \Gamma \mu_S \nabla_\tau (\nabla_\tau \cdot \mathbf{v}) - \rho\nu(\nabla_\tau v_n + \nabla_n \mathbf{v}_\tau) \tag{7b}$$
$$\Gamma \partial_t v_n = \sigma\Delta_\tau \zeta + \Gamma \chi_S \Delta_\tau v_n + p - 2\rho\nu\nabla_n v_n \tag{7c}$$
$$\partial_t \zeta = v_n \tag{7d}$$

The main feature of these balances is the absence of the surface diffusion which is confirmed experimentally [12, 13]. As is seen, eqs. (7) are a particular case of eqs. (6) with $\pi = -\sigma$, $\mathbf{u} = \mathbf{v}$ and $\Gamma_{tot} = \Gamma$. This means that the surface phase should coincide again with the tension, kinematic and equimolecular dividing surfaces. The result that the mass excess of the solvent is zero shows that the insoluble surfactants do not change the place of the surface phase of the pure solvent. They only concentrate there and thus indicate the place of the solvent surface phase. Note that the lack of the surface diffusion is due to the fact that the surface phase is one component, and not because $D_S = 0$. The usual boundary balances of soluble surfactants are [5]

$$\partial_t \Gamma + \Gamma \nabla_\tau \cdot \mathbf{v} = D_S \Delta_\tau \Gamma - D\nabla_n c \tag{8a}$$
$$\partial_\Gamma \sigma \nabla_\tau \Gamma = \rho\nu(\nabla_\tau v_n + \nabla_n \mathbf{v}_\tau) \tag{8b}$$
$$\sigma\Delta_\tau \zeta + p - 2\rho\nu\nabla_n v_n = 0 \tag{8c}$$
$$\partial_t \zeta = v_n \tag{8d}$$

In contrast to the previous case, they depend substantially by the surface diffusion but the effect of the surface viscosity friction is missing. This is in accordance with the experimental results [14, 15]. As is seen, eqs. (8) follow from the general balances (6) when $\pi = -\sigma$, $\mathbf{u} = \mathbf{v}$ and $\Gamma_{tot} = 0$. Therefore, the surface phase of soluble surfactants coincides with the tension, kinematic and zero total mass excess surfaces. This means that the solutions have their own location of the surface phase different from that of the pure solvent. As in the case with the bulk, the surface dynamic viscosities in eq. (4c) are presented as products of the corresponding kinematic viscosities and $\Gamma_{tot}$. For this reason they are absent in the case of soluble surfactants, where $\Gamma_{tot} = 0$. The two distinguished cases discussed above demonstrate limit kinds of relaxation behavior of the surface processes. In general, the surface rheology is described by eqs. (6) not by eqs. (7) or (8). Moreover, the Gibbs approach to interfacial rheology is applicable for fluids with molecules smaller than the transition depth of the interfacial region [4]. For this reason, the presented theory maybe does not work for macromolecular surfactants. In the latter case, additional complication arises from usually non-Newtonian surface rheology of the second component [13].

In the following part the interfacial surface waves for both insoluble and soluble surfactants are studied. As known, they are subject of the light scattering measurements [16]. For the sake of simplicity a planar dividing surface is chosen and one direction propagating waves ($v_y = 0$) are considered. In the frame of the classical hydrodynamics [5], the solutions of eqs. (5) can be presented as

$$v_x = [ikA\exp(kz) - mB\exp(mz)]\exp(ikx + i\omega t) \tag{9a}$$

$$v_z = [kA\exp(kz) + ikB\exp(mz)]\exp(ikx + i\omega t) \tag{9b}$$

$$p = -i\omega\rho A\exp(kz)\exp(ikx + i\omega t) \tag{9c}$$

$$c = C\exp(lz)\exp(ikx + i\omega t) \tag{9d}$$

where $A$, $B$ and $C$ are unknown functions of the wave number $k$ and frequency $\omega$. Here $m^2 \equiv k^2 + i\omega/\nu$ and $l^2 \equiv k^2 + i\omega/D$. For simplicity of the final result, the well-known high viscosity approximation $\omega \ll \nu k^2$ is further applied. Substituting eqs. (9) in eqs. (7) provides the dispersion relations for the case of insoluble surfactants

$$-(\rho + \Gamma k)\omega^2 + 2i\omega\rho\nu k^2 + i\omega\Gamma\chi_s k^3 + \sigma k^3 = 0 \tag{10a}$$

$$-\Gamma k\omega^2 + 2i\omega\rho\nu k^2 + i\omega\Gamma(\nu_s + \mu_s)k^3 + \varepsilon k^3 = 0 \tag{10b}$$

where $\varepsilon = -\Gamma\partial_\Gamma\sigma$ is the surface elasticity.

Usually, the adsorption dynamics of soluble surfactants is presumed to be much quicker than the others relaxation processes. For this reason the equilibrium adsorption isotherm $\Gamma = \Gamma(\rho, c, T)$ will be employed. For the case of soluble surfactants, the combination of eqs. (8) and (9) provides the following dispersion relations

$$-\rho\omega^2 + 2i\omega\rho\nu k^2 + \sigma k^3 = 0 \tag{11a}$$

$$2i\omega\rho\nu k^2 + 2\rho\nu(D_S k^2 + \alpha Dl)k^2 + \varepsilon k^3 = 0 \tag{11b}$$

where $\alpha = \partial_\Gamma c$. Due to the high viscosity approximation, the motion of the interface splits into two independent modes, capillary and longitudinal waves [17]. The common difference between eqs. (10) and (11) is the absence in the latter of the specific surface inertial and viscous terms. This fact is due to the zero total mass density excess at the interfaces of soluble surfactants. In eq. (11b) in addition to (10b) there is a term accounting for the diffusion. As seen, the dispersion relations of the insoluble and soluble surfactants are quite different and provide a good background for verification of the present theory.

1. D. Ronis, D. Bedeaux, I. Oppenheim, *Physica A* **90** (1978) 487
2. S. Ono, S. Kondo, *Molecular Theory of Surface Tension in Liquids*, Springer, Berlin, 1960
3. B. Zielinska, D. Bedeaux, *Physica A* **112** (1982) 265
4. H. Brenner, L. Leal, *J. Colloid Interface Sci.* **88** (1982) 136
5. V. Levich, *Physicochemical Hydrodynamics*, Prentice Hall, Englewood Cliffs, 1962
6. T. Sørensen, *Lect. Notes Phys.* **105** (1979) 1
7. B. Radoev, K. Dimitrov, *Z. Phys. Chem. (Leipzig)* **266** (1985) 1016
8. P. Kralchevsky, I. Ivanov, A. Dimitrov, *Chem. Phys. Lett.* **187** (1991) 129
9. L. Landau, E. Lifshitz, *Fluid Mechanics*, Pergamon, Oxford, 1984
10. Y. Podstrigach, Y. Povstenko, *Introduction to Mechanics of the Surface Phenomena in Deformable Solids*, Naukova Dumka, Kiev, 1985
11. L.E. Scriven, *Chem. Eng. Sci.* **12** (1960) 98
12. D. Dimitrov, I. Panaiotov, P. Richmond, L. Sagara, *J. Colloid Interface Sci.* **65** (1978) 483
13. I. Panaiotov, *Kinetic Properties and State of Insoluble and Soluble Monolayers as Models of Biomembranes*, D.Sc. Thesis, University of Sofia, Sofia, 1986
14. C.S. Vassiliev, E.D. Manev, I.B. Ivanov, *Abhand. Akad. Wiss. DDR, Abtl. Math. Naturwiss. Technik* **1** (1987) 465
15. I. Panaiotov, D. Dimitrov, M. Ivanova, *J. Colloid Interface Sci.* **69** (1979) 318
16. D. Langevin, *Colloids Surf.* **43** (1990) 121
17. R. Tsekov, *Fluctuation Phenomena on Fluid Interfaces,* Ph.D. Thesis, University of Sofia, Sofia, 1993